\documentclass[twocolumn]{aastex62}
\usepackage{amsmath}
\graphicspath{{./}{figures/}}

\received{???}
\revised{???}
\accepted{???}

\usepackage{color}

\shorttitle{parameter free FoG}
\shortauthors{Zhang et al.}

\begin{document}

\title{The parameter-free Finger-Of-God model and its application to 21cm intensity mapping}

\correspondingauthor{Jiajun Zhang}
\email{liambx@ibs.re.kr}
\author[0000-0002-4117-343X]{Jiajun Zhang}
\affil{Center for Theoretical Physics of the Universe, Institute for Basic Science (IBS), Daejeon 34126, Korea}

\author{Andr\'e A. Costa}
\affil{Center for Gravitation and Cosmology, College of Physical Science and Technology, Yangzhou University, Yangzhou 225009, China }

\author{Bin Wang}
\affil{Center for Gravitation and Cosmology, College of Physical Science and Technology, Yangzhou University, Yangzhou 225009, China }

\author{Jianhua He}
\affil{School of Astronomy and Space Science, Nanjing University, Nanjing 210093, China }

\author{Yu Luo}
\affil{Purple Mountain Observatory, Chinese Academy of Sciences, No 10 Yuanhua Road, Nanjing 210033, China}

\author{Xiaohu Yang}
\affil{Department of Astronomy, School of Physics and Astronomy, and Shanghai Key Laboratory for Particle Physics and Cosmology, Shanghai Jiao Tong University, Shanghai 200240, China}
\affil{Tsung-Dao Lee Institute, Shanghai Jiao Tong University, Shanghai 200240, China}


\begin{abstract}

Using the galaxy catalog built from ELUCID N-body simulation and the semi-analytical galaxy formation model, we have built a mock HI intensity mapping map. We have implemented the Finger-of-God (FoG) effect in the map by considering the galaxy HI gas velocity dispersion. By comparing the HI power spectrum in the redshift space with the measurement from IllustrisTNG simulation, we have found that such FoG effect can explain the discrepancy between current mock map built from N-body simulation and Illustris TNG simulation. Then we built a parameter-free FoG model and a shot-noise model to calculate the HI power spectrum. We found that our model can accurately fit both the monopole and quadrupole moments of the HI matter power spectrum. Our method of building the mock HI intensity map and the parameter-free FoG model will be widely useful for the up-coming 21cm intensity mapping experiments, such as CHIME, Tianlai, BINGO, FAST and SKA. It is also crucial for us to study the non-linear effects in 21cm intensity mapping.

\end{abstract}



\section{Introduction} \label{sec:intro}

The 21cm emission line comes from the spin-flip of the electrons in neutral hydrogen. Therefore, the intensity distribution
of 21cm in the universe represents the distribution of neutral hydrogen. The neutral hydrogen traces the underlying matter field, which is also known as the large scale structure of the universe. Just like galaxy surveys, 21cm intensity mapping can also be used to measure the tomographic Baryon Acoustic Oscillations, but much cheaper and faster\citep{Wyithe2008MNRAS}. From upcoming 
experiments such as CHIME\citep{chime}, Tianlai\citep{tianlai}, BINGO\citep{bingo}, FAST\citep{fast2,fast1,fast3}, SKA\citep{ska} and so on, 21cm intensity mapping may be able to tell us the properties of dark matter and dark energy\citep{statusreport}.

As we can only observe the redshifted emission lines of neutral hydrogen rather than their true distance from us, 
the distribution of neutral hydrogen that we can map from the 21cm intensity is distorted by their 
peculiar velocity. This is well known as Redshift-Space Distortions (RSD). Since the peculiar velocity of galaxies are dominated by gravity and the clustering of matter, it contains information of the growth of large scale structure. The RSD effect in galaxy spectroscopy surveys is used to measure the growth factor of the matter power spectrum\citep{sdss4rsd}, test General Relativity\citep{Jullo2019A&A,fT2019PRD} and other cosmological models\citep{Costa2017JCAP,An2019MNRAS,Yang2019PRD,Cheng2019}.

The RSD effect can be understood as a combination of two effects, Kaiser effect\citep{kaiser1987} and Finger-of-God (FoG) effect\citep{FoG}. The Kaiser effect is dominant at large scales, which squeezes the distribution of galaxies in the line-of-sight direction. The FoG effect is dominant in small scales, which elongates the distribution of galaxies, making them look like fingers pointing at the observer. The RSD effect in the 21cm intensity mapping is well studied in \citet{Sarkar19}. They propose a model for calculating the 2D redshift space HI power spectrum , which contains a free parameter $\sigma_p$ for the FoG effect at every different redshift. This free parameter limits our ability to constrain cosmological parameters more precisely. It would be better if we can find a parameter-free model for the FoG effect.

In order to study the RSD effect in 21cm intensity mapping and build a better RSD model, especially at small scales, we need to construct a HI distribution map from high-resolution simulations. We choose to use the ELUCID simulation\citep{ELUCID3} together with a semi-analytical model\citep{Luoyu2016} to build a galaxy catalog. Due to the reionization happened at $z \sim 10$, the universe at $z<5$ is mostly ionized\citep{Becker2001AJ,Fan2006AJ,Fan2006ARA&A}. The only region that neutral hydrogen can "survive" is high density regions\citep{Prochaska2005ApJ,Wolfe2005ARA&A,Zafar2013A&A}, where the column depth can be sufficient for self-shielding\citep{Pritchard2012}. Therefore, at low redshift ($z<5$), most HI gas is inside galaxies, in the form of individual HI clouds. This is also confirmed in the state-of-art hydrodynamic simulation IllustrisTNG\citep{F18}. 

The collective redshifted 21cm emission from these HI clouds appears as a smooth background signal in the radio surveys. Comparing to the resolution of the 21cm intensity mapping surveys, which is square degree level, the size of dark matter halos is negligible. In other words, the galaxies, the dark matter halos and the HI gas inside them can be safely considered as point mass. By smoothing the collective HI mass in the redshift space, we can get a mock HI map. In \citet{HALL2013PRD}, it has been pointed out that the leading contribution to the perturbation of the 21cm intensity (represented in the brightness temperature $T_b$) are the HI density perturbation and redshift-space distortions. Thus, the brightness temperature ($T_b$) fluctuation map of 21cm emission is almost equivalent to the HI density fluctuation map.   

In the next sections, we will first introduce the method that we generate the HI density map from simulations in Sec.\ref{sec:mock}. Then, we will introduce our RSD model and especially the parameter-free FoG model in Sec.\ref{sec:rsd}. We will introduce the model for shot-noise power spectrum in Sec.\ref{sec:noise}. Finally, we will summarize and give some discussions in Sec.\ref{sec:summary}.

\section{Mock HI map building methodology}\label{sec:mock}
\begin{figure*}
    \centering
    \includegraphics[width=\textwidth]{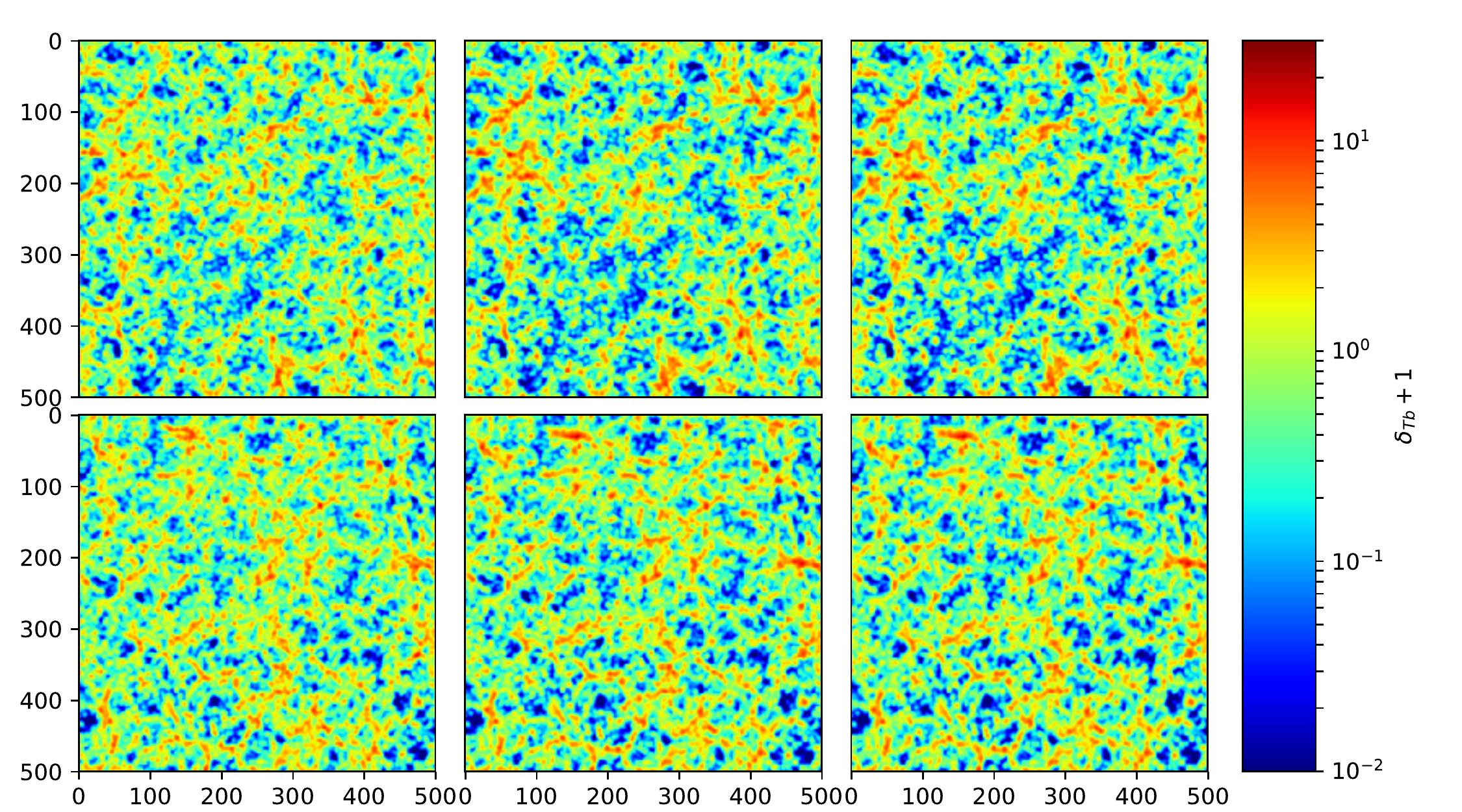}
    \caption{Here is the brightness temperature fluctuation map at $z=0$, the units of axis is $h^{-1}$Mpc. The upper panels are face-on maps with the same slice position and the lower panels are line-of-sight maps with the same slice position. The maps are smoothed by a Gaussian kernel with $\sigma=1h^{-1}$Mpc. From left to right, the panels are in real space, redshift space without velocity dispersion considered and redshift space with velocity dispersion considered. By comparing the upper panels, we can see that for the same slice, which means the same frequency band in observation, the structures we can observe are different in real and redshift spaces. By comparing the lower panels, we can see the squeezing effect, which is the Kaiser effect in the high density regions, and the FoG effect is too tiny to be identified by eye.}
    \label{fig:map}
\end{figure*}
\subsection{Galaxy Catalog}
The galaxy catalog was generated using the ELUCID N-body simulation\citep{ELUCID3}, with the semi-analytical galaxy formation model described in \citet{Luoyu2016}. The ELUCID simulation is run with $3072^3$ particles in a periodic cubic box of $500h^{-1}$Mpc on a side. WMAP5 cosmology\citep{wmap5} was assumed in the ELUCID simulation.
In the galaxy catalog, we use the position, velocity, HI mass and dark matter halo mass of the galaxies, the lower limit of the dark matter halo mass is about $1.85\times10^9 M_{\odot}/h$. We assume that all the HI mass in the universe is inside galaxies and their hosting halos, concentrated in the center of the dark matter halo. As described in \citet{F18} (F18 in short here after), from IllustrisTNG simulation\citep{tng1,tng2,tng3,tng4,tng5} results, it was shown that at $z<3.0$, more than $90\%$ of the HI gas is inside the galaxies and more than $95\%$ of the HI mass is inside the halos. Therefore, our assumption is reasonable.

The HI gas velocity dispersion inside each galaxy, especially inside high mass galaxies, is also important for incorporating the FoG effect into the mock 21cm map\citep{Sarkar19}. We adopted the empirical fitting function provided in F18,
\begin{equation}\label{eq:sigmav}
\sigma_v (M)=\sigma_{10}(\frac{M}{10^{10} h^{-1} M_{\odot}})^{\alpha},
\end{equation}
where $\sigma_{10}$ is in the unit of km/s, $M$ is the dark matter halo mass, $\sigma_{10}=4.8857z+29.95238$ and $\alpha=0.00914286z+0.35714286$ ($z$ means redshift) are fitted from the data provided in F18.

We can construct the HI distribution in both real and redshift spaces from the galaxy catalog, with the HI mass distribution in real space, the bulk velocity of HI mass and the internal velocity dispersion of HI gas.

\subsection{Redshift Space Distortion}
The mapping between the real space position of the HI gas and the redshift space can be separated into two parts, the bulk motion contribution and the internal HI velocity dispersion contribution. The bulk motion contribution can be calculated with the plane-parallel approximation as,
\begin{equation}
\vec{s}=\vec{x}+\frac{1+z}{H(z)}\vec{v}_{\parallel}(\vec{r}),
\end{equation}
where $\vec{v}_{\parallel}(\vec{r})$ is the peculiar velocity of the galaxy (and its HI gas) along the line-of-sight.
In the mock galaxy catalog, all the galaxies and their HI gas inside are simplified as point mass. After moving the positions of the galaxies in real space to redshift space according to their peculiar velocity, they are still point mass. 
However, the internal galaxy HI gas velocity dispersion will stretch the galaxies into needles along the line-of-sight. The amount of stretch can be described by the velocity dispersion,
\begin{equation}
\delta s(M)=\frac{1+z}{H(z)}\frac{\sigma_v (M)}{\sqrt{3}},
\end{equation}
where $M$ is the galaxy dark matter halo mass. Therefore, every galaxy will be stretched differently, according to their halo mass. Since we have assumed that the internal HI velocity distribution is following Gaussian distribution and described by $\sigma_v(M)$, thus the mass profile in the redshift space is also following Gaussian distribution.   
\subsection{Making 3D Map}
We cannot identify a single galaxy in 21cm intensity mapping, we can only observe the collective intensity of the 21cm emission. Thus, we should make 21cm intensity map with grids. For simplicity, we generated 3D brightness temperature fluctuation map from the mock galaxy distribution box. For making 3D map in real space and redshift space without internal velocity dispersion considered, we can first smooth the HI mass into density distribution by nearest-grid-point (NGP) algorithm. For better illustration, we further smooth the HI density field by a Gaussian kernel with $\sigma=1h^{-1}$Mpc, we, however, did not do such smoothing for measuring the power spectrum from the mock map. The HI density is related to the brightness temperature by,
\begin{equation}\label{eq:Tb}
T_b(\vec{x})=189h(\frac{H_0(1+z)^2}{H(z)})\frac{\rho_{\textit{HI}}(\vec{x})}{\rho_{crit}}\textit{mK}.
\end{equation}
It should be noticed that Eq.\ref{eq:Tb} is only applied for the average brightness temperature. The fluctuation of the brightness temperature contains additional contributions, which is well introduced in \citet{HALL2013PRD}. However, since the leading contributions are the HI density distribution and redshift-space distortions, we assume that the residual contributions can be neglected. This assumption is also adopted by F18 to generate the mock map. Under this assumption, the brightness temperature fluctuation can be easily calculated from,
\begin{equation}
\delta_{T_b(\vec{x})} = \frac{T_b(\vec{x})-\bar{T_b}}{\bar{T_b}} = \delta \rho_{\textit{HI}}(\vec{x}) = \frac{\rho_{\textit{HI}}(\vec{x}) - \bar{\rho_{\textit{HI}}}}{\bar{\rho_{\textit{HI}}}}.
\end{equation}

In order to make the 3D brightness temperature fluctuation map with the internal HI velocity dispersion, which is more realistic, here are the steps.
\begin{itemize}
\item[1] Slice the $500h^{-1}$Mpc box into $500^3$ grids, each grid is $1h^{-1}$Mpc in a side.
\item[2] Label the slice in the line-of-sight direction.
\item[3] For the $n^{th}$ slice, pick out the galaxies in the mock catalog within $3\delta s(M)$ away from the slice. The periodic boundary condition should be taken into account properly.
\item[4] Calculate the HI mass weighted contribution from every galaxy picked out. Since we assume the velocity distribution of HI in each galaxy follows a Gaussian distribution, the HI mass distribution in redshift space along the line-of-sight also follows a Gaussian distribution. Thus, the HI mass weight from each galaxy to the $n^{th}$ slice can be calculated by the error function $w=erf(\frac{z-z(n+1)}{\delta s(M)})-erf(\frac{z-z(n)}{\delta s(M)})$, where $z$ is the line-of-sight redshift space position of the galaxy, $z(n)$ and $z(n+1)$ are the edges of the slice. The mass contribution of the galaxy is thus given by $M\times w$
\item[5] Smooth the mass in the $n^{th}$ slice into grids by NGP algorithm.
\item[6] Loop through all $500$ slices, we get the final 3D HI density map, and translate into the brightness temperature fluctuation map.  
\end{itemize}

In Fig.\ref{fig:map}, we have shown the mock brightness temperature fluctuation map in real space and redshift space, in both x-y direction (perpendicular to the line-of-sight) and x-z direction (parallel to the line-of-sight). The FoG effect contributed from the internal HI velocity dispersion is hard to tell by comparing the lower-middle panel and the lower-right panel. The Kaiser effect can be easily seen by comparing the lower-left panel and the lower-middle panel, especially at high density regions such as location around $(x=150,z=20)$ and $(x=450,z=200)$. 
\begin{figure}
	\includegraphics[width=0.5\textwidth]{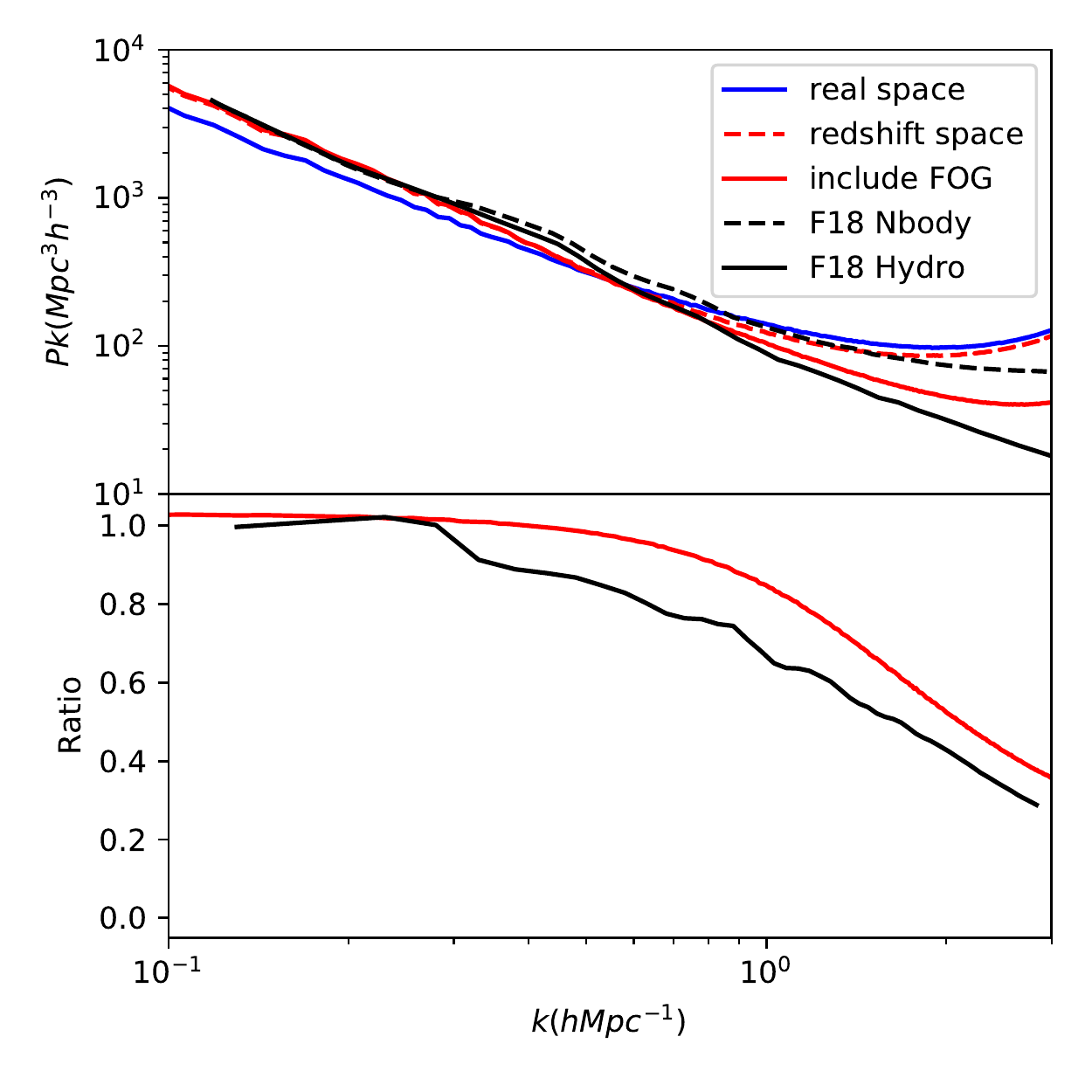}
	\caption{The power spectra measured from our mock map are given in blue solid line (mock in real space), red dashed line (mock in redshift space without FoG) and red solid line (mock in redshift space with FoG). The results from F18 are given by the black dashed line (similar method as our red dashed line) and black solid line (measured from hydrodynamic simulation). In the lower panel, We showed the ratio between solid line and dashed line in the upper panel. In the lower panel, we can see that the difference between red line and black line is very small at large scale and slightly larger but still within $10\%$ at small scale. It proved that our method to generate mock 21cm intensity map, taking internal HI velocity dispersion, is quite successful.}
	\label{fig:pk}
\end{figure}
The power spectrum of our mock map was shown in Fig.\ref{fig:pk}. In F18, the authors have used the N-body simulation together with HI-halo mass relation to assign HI mass into dark matter halos as point mass, which is similar to what we have done here. Without taking the internal HI velocity dispersion into account, they have found that the power spectrum in redshift space measured from hydrodynamic simulation is significantly lower than that of their mock map using N-body simulation. This can be seen in Fig.\ref{fig:pk}, comparing the black dashed line and the black solid line. They have claimed that such suppression at small scale is due to the FoG effect caused by internal velocity dispersion. In Fig.\ref{fig:pk}, comparing the red dashed line and the blue line, we can see that the Kaiser effect can increase the power at large scale, thus tilt the curve more. The ratio between the black solid line and the black dashed line, and ratio between the red solid line and the red dashed line, in the upper panel of Fig.\ref{fig:pk}, are shown in the lower panel. The only difference between red solid line and red dashed line is whether the internal HI velocity dispersion is taken into account. If the claim raised in F18 is correct, that the internal HI velocity dispersion is the reason for such suppression in small scale, the red line representing the ratio will be very close to the black line. We can clearly see that such suppression is really similar as shown in the lower panel of Fig.\ref{fig:pk}.

The red dashed line is similar to the black dashed line at small scales, where they all become flat . The same flatten or even tilted up curve is also seen in blue and red solid lines. This is likely due to the shot noise, since we have assumed that HI mass is all concentrated in the center of the galaxy. This phenomena is not shown in the F18 Hydro curve since the HI mass distribution in the Illustris simulation is better represented with much less shot noise. The shot noise is also part of the reason why the suppression ratio shown as red and black line is still different by about $10\%$. We will discuss the shot noise effect in more detail in Sec.\ref{sec:noise}.

Thus, we think the claim made in F18 that the additional FoG effect caused by the internal HI velocity dispersion can explain the discrepancy between the black solid and dashed lines is correct. After taking the additional FoG effect caused by the HI velocity dispersion inside halos, the discrepancy between the mock and the hydrodynamic simulation result can be resolved to about $10\%$, which is likely due to shot noise. Our method for making mock 21cm intensity map is better than the traditional ways which ignore the internal HI velocity dispersion. It is helpful for our further study about non-linear effect in 21cm intensity mapping.

\section{Redshift Space Distortion Model}\label{sec:rsd}

\begin{figure*}\label{fig:model}
	\includegraphics[width=0.5\textwidth]{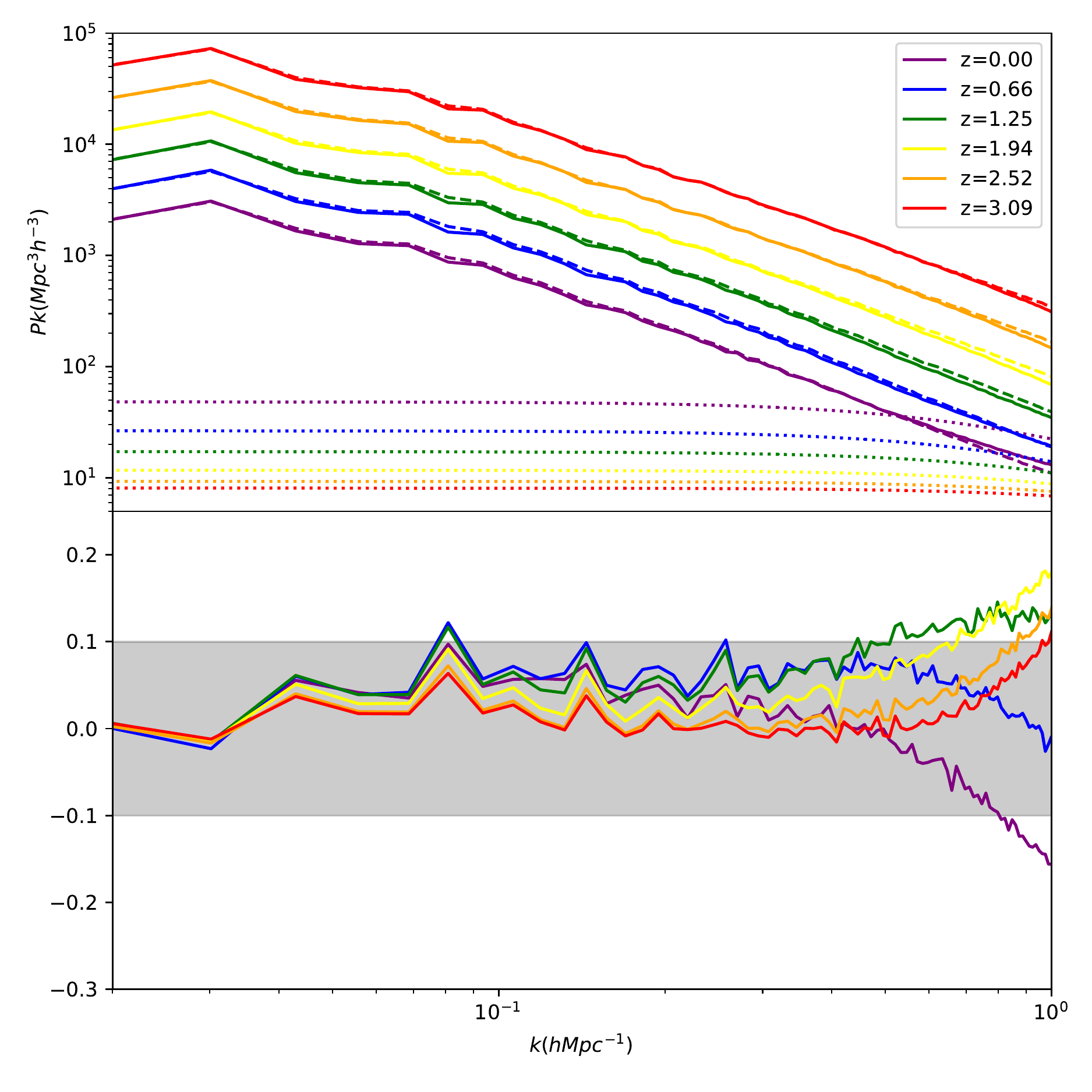}
	\includegraphics[width=0.5\textwidth]{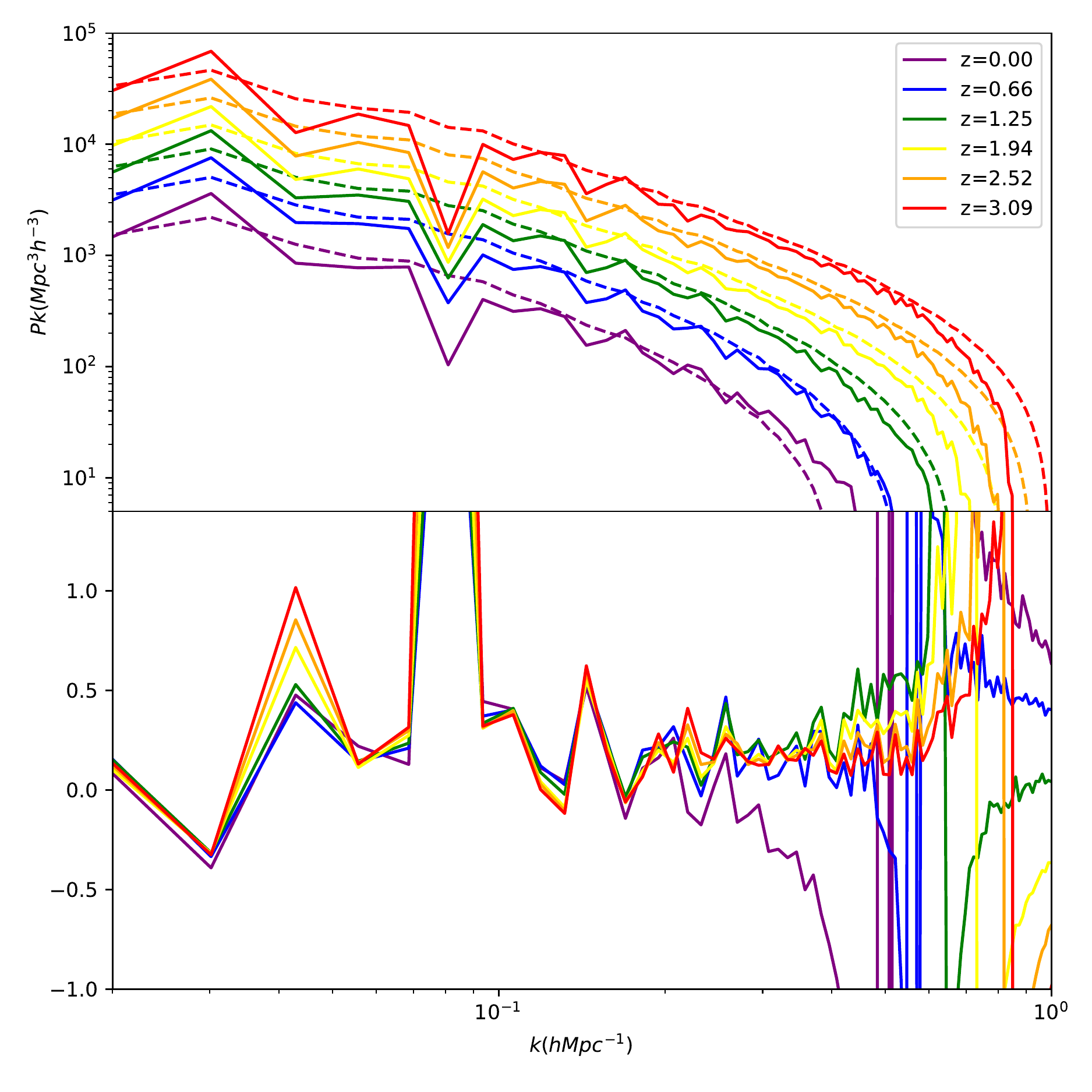}
	\caption{On the left (right) panel, we show the power spectrum monopole $P_{HI}^0$ (quadrupole $P_{HI}^2$). In the upper panels, the measurement from the mock map is given in solid lines and the calculation from our RSD model is shown in dashed lines. Different colors represent different redshifts. The lines in the upper panels are artificially shifted by factor of 4 (2, 1, 0.5, 0.25 and 0.125) for $z=3.09$ (2.52, 1.94, 1.25, 0.66 and 0), for better illustration. In the lower panels, the relative difference ($P_{model}/P_{mock}-1$) between the model calculation and measurement is shown. The shot noise contribution is shown in dotted lines, notice that they have not been artificially shifted. Our model is accurate to $10\%$ for the monopole calculation down to $k=0.4h\mathrm{Mpc}^{-1}$ for all redshifts, and accurate to $20\%$ down to $k=1.0h\mathrm{Mpc}^{-1}$. The model calculation of the quadrupole is also reasonably well considering that the measurement of mock is also highly noisy. }
\end{figure*}

In this section, we would like to introduce the model to calculate the redshift space power spectrum and the comparison with our mock map. The previous efforts as in \citet{F18} and \citet{Sarkar19} have proposed RSD models to calculate the HI power spectrum in redshift space. \citet{Sarkar19} built several mock maps and tried to get the model by fitting to the power spectrum measured from the mock maps. We improved this model with less parameters and good precision.
We follow the model used in \citet{Sarkar19},
\begin{equation}
P_{HI}^s(k,\mu)=(1+\beta\mu^2)^2P_{HI}^r(k)D_{FoG}(k,\mu,\sigma_P),
\end{equation}
where $(1+\beta\mu^2)^2$ is the term representing the Kaiser effect and $\beta$ is set to be a free parameter when fitting the model to the mock map. We choose the best fitted functional form for the FoG term $D_{FoG}(k,\mu,\sigma_P)=(1+\frac{1}{2}k^2\mu^2\sigma_P^2)^{-2}$, where $\sigma_P$ is also used in \citet{Sarkar19} as a free parameter to fit the mock map. 

However, from the steps that we generate the mock 3D map, we note that the FoG effect is closely related to the internal HI velocity dispersion $\sigma_v(M)$. Therefore, it is natural to investigate the relation between $\sigma_P$ and $\sigma_v(M)$.
We found that $\sigma_P$ is a weighted average of $\sigma_v(M)$,
\begin{equation}\label{eq:rsdmodel}
\sigma_P=\frac{\int_{0}^{\infty}\sigma_v(M)\Theta(2\frac{1+z}{H(z)}\sigma_v(M)-1)\frac{dn}{dM}dM}{\int_{0}^{\infty}\Theta(2\frac{1+z}{H(z)}\sigma_v(M)-1)\frac{dn}{dM}dM}\frac{1+z}{H(z)},
\end{equation}
where $\Theta$ function represents the Heaviside step function, $\frac{dn}{dM}$ is the halo mass function, which can be calculated following \citet{Tinker08}. We use yt python package to calculate the halo mass function\citep{yt}. 

The physical meaning of Eq.~\ref{eq:rsdmodel} is the weighted average of velocity dispersion of each halo. The weight is given by the HI velocity dispersion of the halo, or in other words, the halo mass. Since the minimum grid size in our mock map is $1h^{-1}$Mpc, any FoG effect that "stretch" the galaxy less than the grid size cannot be captured by the mock map. It is also similar in the real observation. The frequency resolution is limited to be about 1MHz\citep{bingo}, which is about 2Mpc$/h$. We may also have to artificially set bins in redshift to collect enough signals in the bins. This bin size can be about 10MHz, which is about 20Mpc$/h$. In our model, we have taken this bin size effect into account. Any "stretch" caused by the HI velocity dispersion less than half of the grid size is not taken into account in the model, since it also cannot be captured because of the limited size of the bin. Therefore, our FoG model has no free parameter if the $\sigma_v (M)$ function can be determined either by simulations or other independent observations. On the other hand, with our FoG model, if set to be free for fitting, we can also get the $\sigma_v (M)$ function by fitting to the real 21cm intensity mapping observations.

In Fig\ref{fig:model}, we provided the comparison between the measured power spectrum in redshift space from our mock map and the calculation from the RSD model. We separate the power spectrum using Legendre polynomials. The monopole is $P_{HI}^0(k)=\frac{1}{2} \int_{-1}^{1}P_{HI}^s(k,\mu)d\mu$, and the quadrupole is $P_{HI}^2(k)=\frac{5}{2} \int_{-1}^{1}\frac{3\mu^2-1}{2} P_{HI}^s(k,\mu)d\mu$. For the monopole, our model prediction reached $10\%$ accuracy for all redshifts down to $k=0.5h\mathrm{Mpc}^{-1}$, and reached $\sim20\%$ accuracy down to $k=1.0h\mathrm{Mpc}^{-1}$. The tilt-up tail in the power spectrum measured from mock maps are due to the shot noise\citep{F18}. The detailed discussion of shot noise is given in Sec.\ref{sec:noise}. The quadrupole measurement is very noisy due to lack of multiple realizations and large enough box. Given the noisy quadrupole measurement from mock maps, the model prediction is reasonably good. 

\begin{figure*}
	\includegraphics[width=0.5\textwidth]{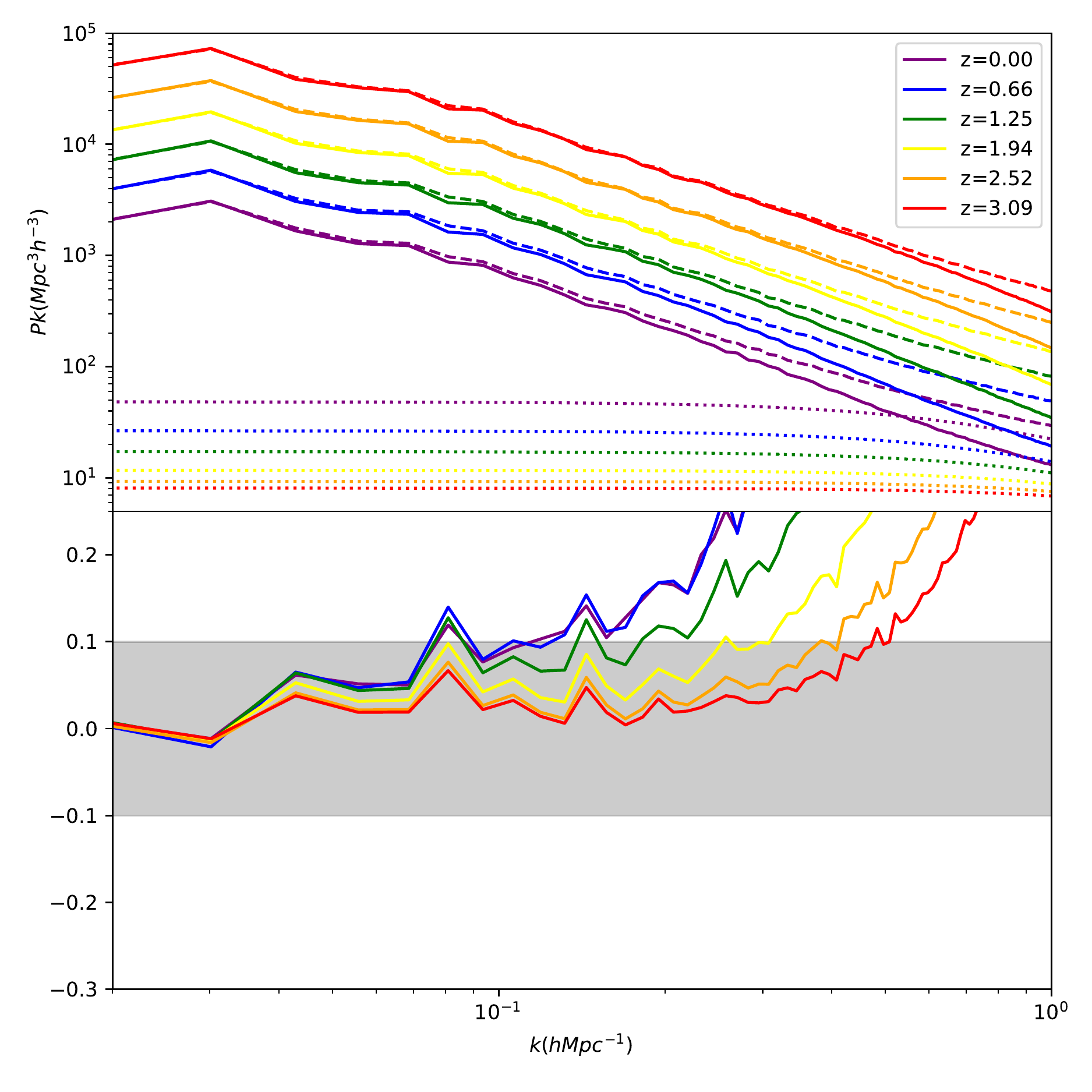}
	\includegraphics[width=0.5\textwidth]{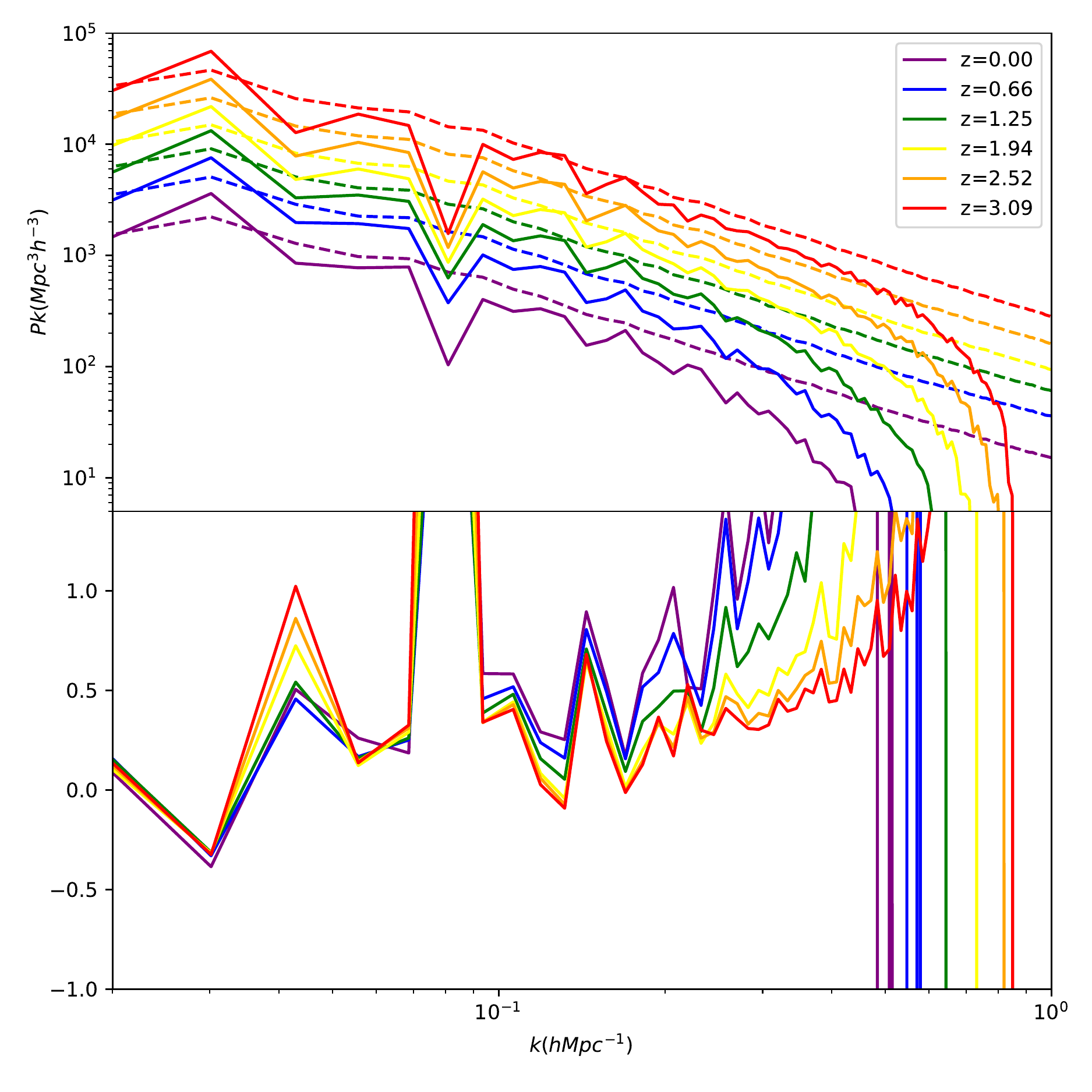}
	\caption{Similar to Fig.\ref{fig:model}, on the left (right) panel, we show the power spectrum monopole $P_{HI}^0$ (quadrupole $P_{HI}^2$). In the upper panels, the measurement from the mock map is given in solid lines and the calculation from our RSD model is shown in dashed lines. Different colors represent different redshifts. The lines in the upper panels are artificially shifted by factor of 4 (2, 1, 0.5, 0.25 and 0.125) for $z=3.09$ (2.52, 1.94, 1.25, 0.66 and 0), for better illustration. The shot noise contribution is shown in dotted lines, notice that they have not been artificially shifted.} In the lower panels, the relative difference ($P_{model}/P_{mock}-1$) between the model calculation and measurement is shown. It is clear that without properly modelling the FoG effect, the calculated HI power spectrum is far from the measurement of the mock.
	\label{fig:nofog}
\end{figure*}

We also calculate the HI power spectrum without considering the FoG effect, the result is shown in Fig.\ref{fig:nofog}. It is quite clear that without modelling the FoG effect, the calculated monopole moment of the HI power spectrum is too high at small scale compared to the mock result. The quadrupole moment of the HI power spectrum is very far from the mock result at small scales. Even at large scales $k>0.1h\textit{Mpc}^{-1}$, the monopole moment of the HI power spectrum calculated from a model without FoG effect is wrong by more than $10\%$ at $z=0.0, 0.66$. This error is larger and larger with increasing $k$. Thus, for a highly accurate observational project aiming to measuring the HI power spectrum in the future, we need a mock which properly takes the FoG effect due to galaxy internal HI velocity dispersion into account, and a model which takes FoG effect into account as well. Our mock building methodology and FoG model provide an important step forward, for the future 21cm Intensity Mapping survey.
\section{Shot Noise}\label{sec:noise}
Since we have put all the HI mass in the center of the halos to generate the mock map, one may naturally raise the question, how will such point mass assumption affect the power spectrum. This question has been well described and discussed in \citet{Castorina2017MNRAS}. The shot noise power spectrum for HI density distribution can be calculated by
\begin{equation}\label{eq:noise}
\begin{aligned}
    P_{HI}^{SN}(z) &= \frac{\int_0^{\infty}n(M,z)M_{HI}^2(M,z)dM}{(\int_0^{\infty}n(M,z)M_{HI}(M,z)dM)^2}\\
                   &= \frac{V}{N}\frac{<M_{HI}^2>}{(<M_{HI}>)^2},
\end{aligned}
\end{equation}
where $V$ is the volume of the simulation box and $N$ is the total number of galaxies in the SAM galaxy catalog which has neutral hydrogen. We have transformed Eq.\ref{eq:noise} to the final form on the right hand side in order to calculate it easier with the catalog, from which we already have the HI mass inside each galaxy. We list the values in Tab.\ref{tab:noise}.
\begin{table}[]
    \centering
    \begin{tabular}{c|c|c|c|c|c|c}
    \hline
        redshift & 0 & 0.66 & 1.25 & 1.94 & 2.52 & 3.09 \\
        \hline
        $\frac{V}{N}/(h^{-1}\mathrm{Mpc})^3$ & 1.82 & 1.80 & 1.82 & 1.94 & 2.13 & 2.44 \\
        \hline
        $\frac{<M_{HI}^2>}{(<M_{HI}>)^2}$ & 26.5 & 14.7 & 9.46 & 6.03 & 4.35 & 3.31\\
        \hline
    \end{tabular}
    \caption{The terms of the shot noise power spectrum calculated from ELUCID SAM galaxy catalog, necessary for calculating the shot noise power spectrum by Eq.\ref{eq:noise} and Eq.\ref{eq:rednoise}.}
    \label{tab:noise}
\end{table}

However, if we would like to calculate the shot noise power spectrum in redshift space, we need to consider the internal HI velocity dispersion. The HI gas in a galaxy is stretched into a line in the line-of-sight direction in redshift space and keeps as point mass in the direction perpendicular to line-of-sight. Therefore, Eq.\ref{eq:noise} is no longer correct. The difference is clear, HI mass is not considered as point mass, but one-dimensional Gaussian density distribution along the line-of-sight direction. So the modified shot noise power spectrum should be
\begin{equation}
    P_{HI}^{sSN}=\frac{\int_0^{\infty}n(M,z)M_{HI}^2(M,z)u(k|\sigma)dM}{(\int_0^{\infty}n(M,z)M_{HI}(M,z)dM)^2},
\end{equation}{}
where
\begin{widetext}
\begin{equation}
    u(k|\sigma)=\frac{\int_{-\infty}^{+\infty}\delta(x)\delta(y)e^{-ik_{x}x}e^{-ik_{y}y}\frac{1}{\sqrt{2\pi\sigma^2}}e^{-z^2/(2\sigma^2)}e^{-ik_{z}z}dxdydz}{\int_{-\infty}^{+\infty}\delta(x)\delta(y)\frac{1}{\sqrt{2\pi\sigma^2}}e^{-z^2/(2\sigma^2)}dxdydz}=e^{-\frac{1}{2}\sigma^2\mu^2 k^2},
\end{equation}{}
\end{widetext}{}
and for simplicity,
\begin{equation}\label{eq:rednoise}
\begin{aligned}
    P_{HI}^{sSN}(k,\mu)&\approx P_{HI}^{SN}u(k|\sigma_P)\\
                       &=\frac{V}{N}\frac{<M_{HI}^2>}{(<M_{HI}>)^2} e^{-\frac{1}{2}\sigma_P^2\mu^2 k^2}.
\end{aligned}
\end{equation}{}
We combine the model for redshift space power spectrum shown in Eq.\ref{eq:rsdmodel} and shot noise contribution in Eq.\ref{eq:rednoise}. The total power spectrum is the sum of Eq.\ref{eq:rsdmodel} and Eq.\ref{eq:rednoise}.
The monopole moment of shot noise power spectrum can be simply calculated by
\begin{equation}
    P_{HI}^{s0SN}=\int_{-1}^{+1}\frac{1}{2}P_{HI}^{sSN}(k,\mu)d\mu,
\end{equation}{}
and the quadrupole moment is
\begin{equation}
    P_{HI}^{s2SN}=\int_{-1}^{+1}\frac{5}{4}(3\mu^2-1)P_{HI}^{sSN}(k,\mu)d\mu.
\end{equation}{}
Since the stretch in redshift space caused by the HI velocity dispersion is quite small, the anisotropy of the shot noise is also very small, the quadrupole moment of the shot noise power spectrum is negligible. For the monopole moment, comparing to the pure point mass shot noise $P_{HI}^{SN}$, $P_{HI}^{s0SN}$ has a smaller value at large $k$. The shot noise contributions are shown as dotted lines in Fig.\ref{fig:model}. The dotted lines have not been artificially shifted, while the solid and dashed lines are shifted, for better illustration. The dotted lines are, in fact always lower than the solid lines and dashed lines. Notice that, in Eq.\ref{eq:noise}, the shot noise power spectrum is independent of $k$, as we expected. It is because of point mass assumption we have made in calculating the shot noise. However, in Eq.\ref{eq:rednoise}, the shot noise power spectrum in redshift space is dependent of $k$ because after considering FoG effect, the point mass assumption is no longer valid. The point mass is replaced by line mass in redshift space. The size of the line is decided by the galaxy HI gas velocity dispersion, therefore the shot noise power spectrum become $k$ dependent.
\begin{figure}
    \centering
    \includegraphics[width=0.5\textwidth]{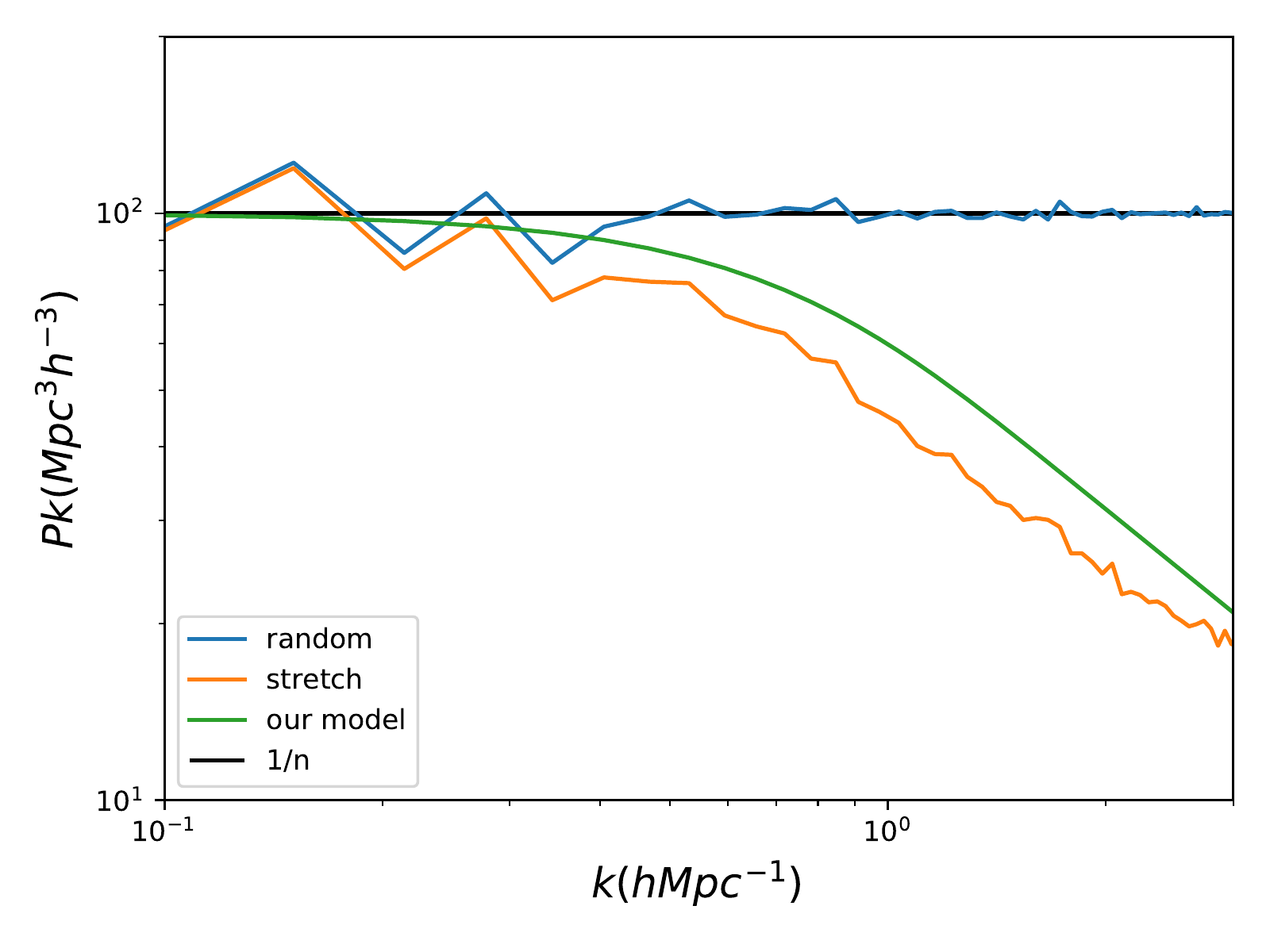}
    \caption{The power spectra of random points is shown in blue. The $1/n$ shot noise is shown in black. The power spectra of stretched random points is shown in orange. The calculated power spectra with Eq.\ref{eq:rednoise} is in green. We can see that our model fits the stretched random sample much better than the traditional $1/n$ shot noise.}
    \label{fig:random}
\end{figure}

In order to justify our scale-dependant shot noise calculation, we have performed a test on random point. We have generated 10000 random points in a $100Mpc/h$ box. Therefore, the number density $n=0.01h^3Mpc^{-3}$. Then we stretched the random points into lines in z direction according to the process for generating the mock catalog, with $\delta_s=5Mpc/h$.  The result is shown in Fig.\ref{fig:random}. We can see that, below Nyquist limit, $1/n$ shot noise can well describe the random point power spectra. However, after stretching, the power spectra drops with smaller scale. Our scale-dependant shot noise model can reasonably describe such power spectra. The random point test shows that it is necessary to take the modification of shot noise power spectrum model into account. Notice that $\delta_s=5Mpc/h$ is not a realistic number, but much larger than the real case. We just use such a large number to enlarge the effect for illustration.
\section{Discussion and Summary}\label{sec:summary}
In this paper, we have provided a novel method to generate mock 21cm intensity map from simulations. This method can take the galaxy internal HI gas velocity dispersion into account correctly. Therefore, it can generate a mock map with fruitful details about Finger-of-God (FoG) effect at small scales. By comparing the power spectrum of our mock map and hydrodynamic simulation shown in F18, we concluded that our method can accurately generate a mock map that includes FoG effect. We have also concluded that the claim raised in F18, that FoG effect can cause additional suppression in the power spectrum, is correct.

We have also proposed a novel RSD model, which includes a parameter-free FoG model. By comparing the power spectrum measured from the mock map and the RSD model, we found that our model can accurately predict the power spectrum measured from the mock maps in redshift space. The parameter-free FoG model is quite useful for future constraints using 21cm intensity mapping. If the HI velocity dispersion versus galaxy virial mass relation can be calibrated by simulations or indenpently measured from other observations, we need one less parameter than the traditional RSD model to fit the power spectrum in redshift space, which will possibly provide better constraints for the cosmological parameters that we are interested in.

However, our mock map still over estimated the power spectrum by about $10\%$, comparing to the result from F18. It is likely due to the low resolution of our mock map ($1h^{-1}$Mpc grid size) and the related shot noise. We also provided a novel calculation for shot noise power spectrum applicable for our mock.

In fact, since most of the near future observations for 21cm intensity mapping is very poor in resolution. The large beam size can effectively suppress the power spectrum at $k>0.2h\mathrm{Mpc}^{-1}$. Therefore, the linear perturbation theory might still be sufficient in the near future. However, if we would like to gain more information and constrain power from the near future surveys, such as CHIME, Tianlai, BINGO, FAST and SKA, a good understanding about the non-linear effects in the small scale is necessary. We have pointed out that the FoG effect due to galaxy internal HI velocity dispersion can lead to an incorrect calculation of HI power spectrum even at $k\sim0.2h\mathrm{Mpc}^{-1}$. Moreover, if we can have higher resolution 21cm intensity mapping surveys in the future, the non-linear effects will be more important. 

\section{Acknowledgement}
Throughout this paper, we have heavily used numpy, scipy and matplotlib for data analysis and making plots\citep{numpy,scipy,matplotlib}. We have also used Pylians\footnote{https://github.com/franciscovillaescusa/Pylians} python package to calculate the power spectrum and density field. This work was supported by IBS under the project code, IBS-R018-D1. Y.Luo acknowledge the support from NSFC (No.11703091). This work was also partially supported by NNSFC key project No:11835009.

\bibliographystyle{aasjournal}
\bibliography{21cm}

\begin{thebibliography}{}
\expandafter\ifx\csname natexlab\endcsname\relax\def\natexlab#1{#1}\fi
\providecommand{\url}[1]{\href{#1}{#1}}

\bibitem[{{An} {et~al.}(2019){An}, {Costa}, {Xiao}, {Zhang}, \&
  {Wang}}]{An2019MNRAS}
{An}, R., {Costa}, A.~A., {Xiao}, L., {Zhang}, J., \& {Wang}, B. 2019, \mnras,
  489, 297

\bibitem[{{Anagnostopoulos} {et~al.}(2019){Anagnostopoulos}, {Basilakos}, \&
  {Saridakis}}]{fT2019PRD}
{Anagnostopoulos}, F.~K., {Basilakos}, S., \& {Saridakis}, E.~N. 2019, \prd,
  100, 083517

\bibitem[{{Becker} {et~al.}(2001){Becker}, {Fan}, {White}, {Strauss},
  {Narayanan}, {Lupton}, {Gunn}, {Annis}, {Bahcall}, {Brinkmann}, {Connolly},
  {Csabai}, {Czarapata}, {Doi}, {Heckman}, {Hennessy}, {Ivezi{\'c}}, {Knapp},
  {Lamb}, {McKay}, {Munn}, {Nash}, {Nichol}, {Pier}, {Richards}, {Schneider},
  {Stoughton}, {Szalay}, {Thakar}, \& {York}}]{Becker2001AJ}
{Becker}, R.~H., {Fan}, X., {White}, R.~L., {et~al.} 2001, \aj, 122, 2850

\bibitem[{{Bigot-Sazy} {et~al.}(2016){Bigot-Sazy}, {Ma}, {Battye}, {Browne},
  {Chen}, {Dickinson}, {Harper}, {Maffei}, {Olivari}, \&
  {Wilkinsondagger}}]{fast2}
{Bigot-Sazy}, M.~A., {Ma}, Y.~Z., {Battye}, R.~A., {et~al.} 2016, Astronomical
  Society of the Pacific Conference Series, Vol. 502, {HI Intensity Mapping
  with FAST}, ed. L.~{Qain} \& D.~{Li}, 41

\bibitem[{{Castorina} \& {Villaescusa-Navarro}(2017)}]{Castorina2017MNRAS}
{Castorina}, E., \& {Villaescusa-Navarro}, F. 2017, \mnras, 471, 1788

\bibitem[{{Chen}(2012)}]{tianlai}
{Chen}, X. 2012, in International Journal of Modern Physics Conference Series,
  Vol.~12, International Journal of Modern Physics Conference Series, 256--263

\bibitem[{{Cheng} {et~al.}(2019){Cheng}, {Ma}, {Wu}, {Zhang}, \&
  {Chen}}]{Cheng2019}
{Cheng}, G., {Ma}, Y., {Wu}, F., {Zhang}, J., \& {Chen}, X. 2019, arXiv
  e-prints, arXiv:1911.04520

\bibitem[{{Costa} {et~al.}(2017){Costa}, {Xu}, {Wang}, \&
  {Abdalla}}]{Costa2017JCAP}
{Costa}, A.~A., {Xu}, X.-D., {Wang}, B., \& {Abdalla}, E. 2017, \jcap, 2017,
  028

\bibitem[{{Dunkley} {et~al.}(2009){Dunkley}, {Komatsu}, {Nolta}, {Spergel},
  {Larson}, {Hinshaw}, {Page}, {Bennett}, {Gold}, {Jarosik}, {Weiland},
  {Halpern}, {Hill}, {Kogut}, {Limon}, {Meyer}, {Tucker}, {Wollack}, \&
  {Wright}}]{wmap5}
{Dunkley}, J., {Komatsu}, E., {Nolta}, M.~R., {et~al.} 2009, \apjs, 180, 306

\bibitem[{{Fan} {et~al.}(2006{\natexlab{a}}){Fan}, {Carilli}, \&
  {Keating}}]{Fan2006ARA&A}
{Fan}, X., {Carilli}, C.~L., \& {Keating}, B. 2006{\natexlab{a}}, \araa, 44,
  415

\bibitem[{{Fan} {et~al.}(2006{\natexlab{b}}){Fan}, {Strauss}, {Becker},
  {White}, {Gunn}, {Knapp}, {Richards}, {Schneider}, {Brinkmann}, \&
  {Fukugita}}]{Fan2006AJ}
{Fan}, X., {Strauss}, M.~A., {Becker}, R.~H., {et~al.} 2006{\natexlab{b}}, \aj,
  132, 117

\bibitem[{{Hall} {et~al.}(2013){Hall}, {Bonvin}, \& {Challinor}}]{HALL2013PRD}
{Hall}, A., {Bonvin}, C., \& {Challinor}, A. 2013, \prd, 87, 064026

\bibitem[{{Hu} {et~al.}(2019){Hu}, {Wang}, {Wu}, {Wang}, {Zhang}, \&
  {Chen}}]{fast3}
{Hu}, W., {Wang}, X., {Wu}, F., {et~al.} 2019, arXiv e-prints, arXiv:1909.10946

\bibitem[{{Hunter}(2007)}]{matplotlib}
{Hunter}, J.~D. 2007, Computing in Science and Engineering, 9, 90

\bibitem[{{Icaza-Lizaola} {et~al.}(2019){Icaza-Lizaola}, {Vargas-Maga{\~n}a},
  {Fromenteau}, {Alam}, {Camacho}, {Gil-Marin}, {Paviot}, {Ross}, {Schneider},
  {Tinker}, {Wang}, {Zhao}, {Prakash}, {Rossi}, {Zao}, {Cruz-Gonzalez}, \& {de
  la Macorra}}]{sdss4rsd}
{Icaza-Lizaola}, M., {Vargas-Maga{\~n}a}, M., {Fromenteau}, S., {et~al.} 2019,
  arXiv e-prints, arXiv:1909.07742

\bibitem[{{Jackson}(1972)}]{FoG}
{Jackson}, J.~C. 1972, \mnras, 156, 1P

\bibitem[{{Jullo} {et~al.}(2019){Jullo}, {de la Torre}, {Cousinou},
  {Escoffier}, {Giocoli}, {Metcalf}, {Comparat}, {Shan}, {Makler}, {Kneib},
  {Prada}, {Yepes}, \& {Gottl{\"o}ber}}]{Jullo2019A&A}
{Jullo}, E., {de la Torre}, S., {Cousinou}, M.~C., {et~al.} 2019, \aap, 627,
  A137

\bibitem[{{Kaiser}(1987)}]{kaiser1987}
{Kaiser}, N. 1987, \mnras, 227, 1

\bibitem[{{Kovetz} {et~al.}(2017){Kovetz}, {Viero}, {Lidz}, {Newburgh},
  {Rahman}, {Switzer}, {Kamionkowski}, {Aguirre}, {Alvarez}, {Bock}, {Bond},
  {Bower}, {Bradford}, {Breysse}, {Bull}, {Chang}, {Cheng}, {Chung}, {Cleary},
  {Corray}, {Crites}, {Croft}, {Dor{\'e}}, {Eastwood}, {Ferrara}, {Fonseca},
  {Jacobs}, {Keating}, {Lagache}, {Lakhlani}, {Liu}, {Moodley}, {Murray},
  {P{\'e}nin}, {Popping}, {Pullen}, {Reichers}, {Saito}, {Saliwanchik},
  {Santos}, {Somerville}, {Stacey}, {Stein}, {Villaescusa-Navarro}, {Visbal},
  {Weltman}, {Wolz}, \& {Zemcov}}]{statusreport}
{Kovetz}, E.~D., {Viero}, M.~P., {Lidz}, A., {et~al.} 2017, arXiv e-prints,
  arXiv:1709.09066

\bibitem[{{Luo} {et~al.}(2016){Luo}, {Kang}, {Kauffmann}, \& {Fu}}]{Luoyu2016}
{Luo}, Y., {Kang}, X., {Kauffmann}, G., \& {Fu}, J. 2016, \mnras, 458, 366

\bibitem[{{Naiman} {et~al.}(2018){Naiman}, {Pillepich}, {Springel},
  {Ramirez-Ruiz}, {Torrey}, {Vogelsberger}, {Pakmor}, {Nelson}, {Marinacci},
  {Hernquist}, {Weinberger}, \& {Genel}}]{tng1}
{Naiman}, J.~P., {Pillepich}, A., {Springel}, V., {et~al.} 2018, \mnras, 477,
  1206

\bibitem[{{Nelson} {et~al.}(2018){Nelson}, {Pillepich}, {Springel},
  {Weinberger}, {Hernquist}, {Pakmor}, {Genel}, {Torrey}, {Vogelsberger},
  {Kauffmann}, {Marinacci}, \& {Naiman}}]{tng2}
{Nelson}, D., {Pillepich}, A., {Springel}, V., {et~al.} 2018, \mnras, 475, 624

\bibitem[{{Nelson} {et~al.}(2019){Nelson}, {Springel}, {Pillepich},
  {Rodriguez-Gomez}, {Torrey}, {Genel}, {Vogelsberger}, {Pakmor}, {Marinacci},
  {Weinberger}, {Kelley}, {Lovell}, {Diemer}, \& {Hernquist}}]{tng5}
{Nelson}, D., {Springel}, V., {Pillepich}, A., {et~al.} 2019, Computational
  Astrophysics and Cosmology, 6, 2

\bibitem[{{Pillepich} {et~al.}(2018){Pillepich}, {Nelson}, {Hernquist},
  {Springel}, {Pakmor}, {Torrey}, {Weinberger}, {Genel}, {Naiman}, {Marinacci},
  \& {Vogelsberger}}]{tng3}
{Pillepich}, A., {Nelson}, D., {Hernquist}, L., {et~al.} 2018, \mnras, 475, 648

\bibitem[{{Pritchard} \& {Loeb}(2012)}]{Pritchard2012}
{Pritchard}, J.~R., \& {Loeb}, A. 2012, Reports on Progress in Physics, 75,
  086901

\bibitem[{{Prochaska} {et~al.}(2005){Prochaska}, {Herbert-Fort}, \&
  {Wolfe}}]{Prochaska2005ApJ}
{Prochaska}, J.~X., {Herbert-Fort}, S., \& {Wolfe}, A.~M. 2005, \apj, 635, 123

\bibitem[{{Santos} {et~al.}(2015){Santos}, {Bull}, {Alonso}, {Camera},
  {Ferreira}, {Bernardi}, {Maartens}, {Viel}, {Villaescusa-Navarro}, {Abdalla},
  {Jarvis}, {Metcalf}, {Pourtsidou}, \& {Wolz}}]{ska}
{Santos}, M., {Bull}, P., {Alonso}, D., {et~al.} 2015, Advancing Astrophysics
  with the Square Kilometre Array (AASKA14), 19

\bibitem[{{Sarkar} \& {Bharadwaj}(2019)}]{Sarkar19}
{Sarkar}, D., \& {Bharadwaj}, S. 2019, \mnras, 487, 5666

\bibitem[{{Smoot} \& {Debono}(2017)}]{fast1}
{Smoot}, G.~F., \& {Debono}, I. 2017, \aap, 597, A136

\bibitem[{{Springel} {et~al.}(2018){Springel}, {Pakmor}, {Pillepich},
  {Weinberger}, {Nelson}, {Hernquist}, {Vogelsberger}, {Genel}, {Torrey},
  {Marinacci}, \& {Naiman}}]{tng4}
{Springel}, V., {Pakmor}, R., {Pillepich}, A., {et~al.} 2018, \mnras, 475, 676

\bibitem[{{Tinker} {et~al.}(2008){Tinker}, {Kravtsov}, {Klypin}, {Abazajian},
  {Warren}, {Yepes}, {Gottl{\"o}ber}, \& {Holz}}]{Tinker08}
{Tinker}, J., {Kravtsov}, A.~V., {Klypin}, A., {et~al.} 2008, \apj, 688, 709

\bibitem[{{Turk} {et~al.}(2011){Turk}, {Smith}, {Oishi}, {Skory}, {Skillman},
  {Abel}, \& {Norman}}]{yt}
{Turk}, M.~J., {Smith}, B.~D., {Oishi}, J.~S., {et~al.} 2011, The Astrophysical
  Journal Supplement Series, 192, 9

\bibitem[{{van der Walt} {et~al.}(2011){van der Walt}, {Colbert}, \&
  {Varoquaux}}]{numpy}
{van der Walt}, S., {Colbert}, S.~C., \& {Varoquaux}, G. 2011, Computing in
  Science and Engineering, 13, 22

\bibitem[{{Vanderlinde} \& {Chime Collaboration}(2014)}]{chime}
{Vanderlinde}, K., \& {Chime Collaboration}. 2014, in Exascale Radio Astronomy,
  Vol.~2

\bibitem[{{Villaescusa-Navarro} {et~al.}(2018){Villaescusa-Navarro}, {Genel},
  {Castorina}, {Obuljen}, {Spergel}, {Hernquist}, {Nelson}, {Carucci},
  {Pillepich}, {Marinacci}, {Diemer}, {Vogelsberger}, {Weinberger}, \&
  {Pakmor}}]{F18}
{Villaescusa-Navarro}, F., {Genel}, S., {Castorina}, E., {et~al.} 2018, \apj,
  866, 135

\bibitem[{{Virtanen} {et~al.}(2019){Virtanen}, {Gommers}, {Oliphant},
  {Haberland}, {Reddy}, {Cournapeau}, {Burovski}, {Peterson}, {Weckesser},
  {Bright}, {van der Walt}, {Brett}, {Wilson}, {Jarrod Millman}, {Mayorov},
  {Nelson}, {Jones}, {Kern}, {Larson}, {Carey}, {Polat}, {Feng}, {Moore},
  {VanderPlas}, {Laxalde}, {Perktold}, {Cimrman}, {Henriksen}, {Quintero},
  {Harris}, {Archibald}, {Ribeiro}, {Pedregosa}, {van Mulbregt}, \&
  {Contributors}}]{scipy}
{Virtanen}, P., {Gommers}, R., {Oliphant}, T.~E., {et~al.} 2019, arXiv
  e-prints, arXiv:1907.10121

\bibitem[{{Wang} {et~al.}(2016){Wang}, {Mo}, {Yang}, {Zhang}, {Shi}, {Jing},
  {Liu}, {Li}, {Kang}, \& {Gao}}]{ELUCID3}
{Wang}, H., {Mo}, H.~J., {Yang}, X., {et~al.} 2016, \apj, 831, 164

\bibitem[{{Wolfe} {et~al.}(2005){Wolfe}, {Gawiser}, \&
  {Prochaska}}]{Wolfe2005ARA&A}
{Wolfe}, A.~M., {Gawiser}, E., \& {Prochaska}, J.~X. 2005, \araa, 43, 861

\bibitem[{{Wuensche} {et~al.}(2019){Wuensche}, {pre=``}, \& {the`` BINGO
  Collaboration}}]{bingo}
{Wuensche}, C.~A., {pre=``}, l., \& {the`` BINGO Collaboration}. 2019, in
  Journal of Physics Conference Series, Vol. 1269, Journal of Physics
  Conference Series, 012002

\bibitem[{{Wyithe} {et~al.}(2008){Wyithe}, {Loeb}, \& {Geil}}]{Wyithe2008MNRAS}
{Wyithe}, J. S.~B., {Loeb}, A., \& {Geil}, P.~M. 2008, \mnras, 383, 1195

\bibitem[{{Yang} {et~al.}(2019){Yang}, {Shahalam}, {Pal}, {Pan}, \&
  {Wang}}]{Yang2019PRD}
{Yang}, W., {Shahalam}, M., {Pal}, B., {Pan}, S., \& {Wang}, A. 2019, \prd,
  100, 023522

\bibitem[{{Zafar} {et~al.}(2013){Zafar}, {P{\'e}roux}, {Popping}, {Milliard},
  {Deharveng}, \& {Frank}}]{Zafar2013A&A}
{Zafar}, T., {P{\'e}roux}, C., {Popping}, A., {et~al.} 2013, \aap, 556, A141

\end{thebibliography}



\end{document}